

\def\|{\kern3pt}

\def\post #1. #2\par{\medbreak
  \noindent{P\|o\|s\|t\|u\|l\|a\|t\|e\kern8pt #1.\enspace}{\sl#2\par}%
  \ifdim\lastskip<\medskipamount \removelastskip\penalty55\medskip\fi}

\def\anti#1{\bar{#1}}
\def\eps{\epsilon}
\def\lra{\longleftrightarrow}
\def\qq{Q^\prime}
\def\dd{\natural}
\centerline{A model of the composite structure of quarks and leptons
with $SU(4)$ gauge symmetry}
\vskip 1cm
\centerline{ N.G.Marchuk }
\vskip 1cm
\centerline{ Abstract }
\medskip

{\sl A model in which quarks and leptons consist of three
"more elementary" particles of spin 1/2 is proposed. A gauge field theory with
$SU(4)$ symmetry that corresponds to this model predicts the
existence of two new bosons.}
\vskip 2cm

{\bf 1. Introduction.} During the last two decades the so-called
Standard Model (SM) has a stable success in the description of elementary
particle physics. The following particles are considered in the SM as
elementary particles: quarks with three colors (red, yellow, blue)
and six flavors - $u,d,c,s,t,b$; leptons -
$e^-, \nu_e, \mu^-, \nu_\mu, \tau^-, \nu_\tau$;
quanta of electroweak interactions - photon $\gamma$ and
$W^\pm, Z^0$ bosons; Higgs particle $H$; quanta of strong interactions -
gluons $g$ and antiparticles to all listed particles.
Quarks and leptons are called fundamental fermions (FF), and quanta
of electroweak and strong interactions are called fundamental bosons
(FB). FF can be divided into three generations
$$
\pmatrix{ \nu_e & u \cr e^- & d}, \quad
\pmatrix{ \nu_\mu & c \cr \mu^- & s}, \quad
\pmatrix{ \nu_\tau & t \cr \tau^- & b}. \eqno(1)
$$
Respective particles in different generations have similar properties
but different masses that grow from first to third generation
(masses of all types of neutrino possibly equal to zero). On that basis
one can suppose an existence of the entire structure of quarks and leptons.

From the beginning of the seventies  several models
have been created
that consider quarks and leptons as composite particles consisting of
"more elementary" particles -
preons, subquarks, maons, alphons,
quinks, rishons, quips, haplons and so on.
Almost all of those models can be included into one of the two classes -
trio-fermion models in which quarks and leptons consist of three preons
with spin 1/2 [1-13] and fermion-boson models in which quarks and
leptons consist of one preon with spin 1/2 and one preon with spin 0
[14-18] (in the list of references only those articles are mentioned
which were accessible to the present
author). Among trio-fermion models, from our point of view, a most
natural one is TCA-model of Terazawa, Chikashige and Akama [7] that
collect several features from predecessors' models of Senju [2,3] and
Pati, Salam and Strathdee [6,14]. Also, TCA-model is universal enough.
As was mentioned by Terazawa [8], almost all models that were suggested
before or after TCA-model can be considered as special cases of
TCA-model. In particular this is true for the model [1] that was
proposed as an attempt to improve Harari-Shupe model [9-12]. The
universality of TCA-model has it adverse side - apart of 24 FF and 12 FB
of SM TCA-model predicts the existence of more than 200 new fermions and
bosons. That number is, probably, too big for the model that pretends to
be realistic. So, if one wants to develop a realistic preon model, then one
has to include in it some additional conditions that restrict a number
of predicted new particles. An interesting attempt in this direction
was made by Senju [4,5],
who suggests that three preons in the particle have new preonic charge
1,1,-2 (he discussed the possibility to use a magnetic charge). In his
resulting model a number of new particles is less than in TCA-model.

In the model [1] a postulate was formulated
that a particle which
consists of three preons cannot contain two preons with the same sign of
electric charge. A further development of this postulate leads us to the
present model in which there are no new fermions and only two new
bosons $X,Y$ (boson $X$ with three colors).
\bigbreak
{\bf A description of the model.} Let us assume the existence of nine
elementary particles with spin 1/2
$$
\alpha,\beta^r,\beta^y,\beta^b,\eps^u,\eps^d,\delta^1,\delta^2,\delta^3
\eqno(2)
$$
Following [1], we shall call them inds\footnote{\dag}{Ind is a river
taking it source in Himalayas. First draft of the model was done in 1993
in Bangalore during monsoon season.}
Superscripts $r,y,b$ indicate colors of inds  $\beta^r,\beta^y,\beta^b$;
superscripts $u,d$ show the properties upness and downness of inds
$\eps^u,\eps^d$ that associate with $u,d$ quarks and with neutrino
and electron; superscripts 1,2,3 of inds $\delta^1,\delta^2,\delta^3$
indicate a generation of a particle. The quantum numbers of inds
are collected in the table 1.
\medskip
Table 1.
$$
\matrix{
\hbox{Ind}&\hbox{Color}&{B}&{L}&{Q}&{Y^w}&{I^w}&{I^w_3}&
\hbox{Mass} \cr
\alpha    & 0       & 0    & 1  & -1/2-\qq & -1-2\qq & 0 & 0 & 0 \cr
\beta    & {r,y,b}  & 1/3    & 0  & 1/6-\qq & 1/3-2\qq & 0 & 0& 0  \cr
\eps^u_L & 0         & 0      & 0  & 1/2+\qq & 2\qq     & 1/2 & 1/2& 0 \cr
\eps^u_R & 0         & 0      & 0  & 1/2+\qq & 1+2\qq     & 0 & 0& 0  \cr
\eps^d_L & 0         & 0      & 0  & -1/2+\qq & 2\qq     & 1/2 & -1/2& 0 \cr
\eps^d_R & 0         & 0      & 0  & -1/2+\qq & -1+2\qq     & 0 & 0& 0  \cr
\delta^1 & 0 & 0 &0 & 0& 0 &0 & 0& m_1\cr
\delta^2 & 0 & 0 &0 & 0& 0 &0 & 0& m_2\cr
\delta^3 & 0 & 0 &0 & 0& 0 &0 & 0& m_3\cr
}
$$
\medskip
Here $B$ - barion number; $L$ - lepton number; $Q$ - electric charge in
units of proton charge; $Y^w$ - weak hypercharge; $I^w$ - weak isospin;
$I^w_3$ - third component of weak isospin. The quantum numbers $Q,Y^w,I^w_3$
connected with each other by the Gell-Mann - Nishijima formula
$Q=I^w_3+Y^w/2$. Subscripts $L,R$ indicate that we consider left-handed
and right-handed particles. The electric charge and the weak hypercharge
of inds depend on unknown value of electric charge $Q^\prime$ which we
will discuss later.

Each FF of the SM (18 quarks and 6 leptons) we identify with a trio of inds as
shown in the table 2 for the first generation particles
\medskip
Table 2.
$$
\matrix{
\hbox{FF}&\hbox{Inds}&\hbox{Color}&{B}&{L}&{Q}&{Y^w}&{I^w}&{I^w_3}\cr
{\nu_e}_L&\alpha_L\eps^u_L\delta^1_R&0&0&1&0&-1&1/2&1/2\cr
{\nu_e}_R&\alpha_R\eps^u_R\delta^1_L&0&0&1&0& 0&  0&  0\cr
e^-_L    &\alpha_L\eps^d_L\delta^1_R&0&0&1&-1&-1&1/2&-1/2\cr
e^-_R    &\alpha_R\eps^d_R\delta^1_L&0&0&1&-1&-2&  0&   0\cr
u_L    &\beta_L\eps^u_L\delta^1_R&{r,y,b}&1/3&0&2/3&1/3&1/2&1/2\cr
u_R    &\beta_R\eps^u_R\delta^1_L&{r,y,b}&1/3&0&2/3&4/3&  0&  0\cr
d_L    &\beta_L\eps^d_L\delta^1_R&{r,y,b}&1/3&0&-1/3&1/3&1/2&-1/2\cr
d_R    &\beta_R\eps^d_R\delta^1_L&{r,y,b}&1/3&0&-1/3&-2/3& 0&0}
$$
\medskip

We get tables for FF of second and third generations by replacing ind
$\delta^1$ with $\delta^2$ and $\delta^3$ respectively. As we see
from the tables 1 and 2, the quantum numbers of quarks and leptons are the
precise sums of quantum numbers of inds that constitute FF. The electric
charges and the weak hypercharges of FF do not depend on $Q^\prime$. So the
value $Q^\prime$ is called a hidden electric charge. We do not consider
it as a free parameter of the model supposing that $Q^\prime$ has a
definite value that satisfies a condition $|\qq|>1/2$. But now we don't
know how to calculate or measure it.

From the nine types of inds one can construct 165 trios of inds.
A question arise -- is it possible to point out a
selection rule which distinguishes 24 trios of inds that we identify
with FF from the set of 165 trios? The answer is yes. For this purpose
we have to consider the signs of electric charge of inds in trios. There
are 10 variants:
$(+++)$,$(++0)$,$(++-)$,$(+00)$,$(+0-)$,$(+--)$,$(000)$,$(00-)$,$(0--)$,$(---)$.
If we assume that $|\qq|>1/2$ and consider trios of inds (2) with only
one set of signs of electric charge $(+0-)$, then we see that there are
precisely 24 such trios namely  $\eps\alpha\delta$ (6 pieces) and
$\eps\beta\delta$ (18 pieces), and they correspond to quarks and leptons
in table 2.

I have to admit that I cannot
understand why this selection rule
works. But it does work!

\post 1. Let us assume that $|\qq|>1/2$. From the nine inds (2) fermions
can be
formed as trios of inds in such a way that a fermion must contain
two inds with nonzero electric charge with opposite signs of charge and
one electrically neutral ind. Charged inds have the same handedness and
a neutral ind has opposite handedness (hence, a general spin of trio is
equal to 1/2\footnote{\dag\dag}{G.'t Hooft [19] argues that cannot
exist particles with spin 3/2 which consist of three
massless fermions.}).\par

The postulate 1 we complete with postulate $2^\prime$ about the structure of
bosons. Later we'll replace postulate $2^\prime$ by a better one.

\post $2^\prime$. Let $|\qq|>1/2$. From the nine inds (2) bosons can be
formed  as pairs ind-antiind (or superpositions of such pairs).
Ind and antiind that constitute boson both have nonzero electric charge.
Signs of electric charge of these ind and antiind are opposite and
handedness is the same.\par

Let us consider bosons that consist of pairs ind-antiind. It is easy to check,
that the following set of pairs exhaust all set of pairs that satisfy
postulate $2^\prime$:
$$
\alpha\anti\alpha,\beta^c\anti\beta^{c^\prime},
\eps^u\anti\eps^u,\eps^d\anti\eps^d,\eps^u\anti\eps^d,\eps^d\anti\eps^u,
\alpha\anti\beta^c, \anti\alpha\beta^c \eqno(3)
$$
where indices $c,c^\prime$ denote colors $r,y,b$.

Let us identify $W^\pm$ bosons with pairs from (3) $W^+=\eps^u\anti\eps^d$,
$W^-=\anti\eps^u\eps^d$;
Gluons $g_1,\ldots,g_8$ identify with six pairs
$\beta^c\anti\beta^{c^\prime}, c\neq c^\prime$ and with two
superpositions of states:
${1\over\sqrt2}(\beta^r\anti\beta^r-\beta^y\anti\beta^y)$,
${1\over\sqrt6}(\beta^r\anti\beta^r+\beta^y\anti\beta^y-2\beta^b\anti\beta^b)$.
Let us denote by  $\beta\anti\beta$ color singlet state
$\beta\anti\beta=
{1\over\sqrt3}(\beta^r\anti\beta^r+\beta^y\anti\beta^y+\beta^b\anti\beta^b)$
and identify photon $\gamma$ and $Z$ boson with the following
superpositions of states:
$$
\eqalign{
\gamma&=\phi_1\alpha\anti\alpha+\phi_2\eps^u\anti\eps^u+\phi_3\eps^d\anti\eps^d+
\phi_4\beta\anti\beta\quad
({\phi_1}^2+{\phi_2}^2+{\phi_3}^2+{\phi_4}^2=1), \cr
Z^0&=\theta_1\alpha\anti\alpha+\theta_2\eps^u\anti\eps^u+\theta_3\eps^d\anti\eps^d+
\theta_4\beta\anti\beta\quad
({\theta_1}^2+{\theta_2}^2+{\theta_3}^2+{\theta_4}^2=1).
}
$$
So, from the set (3) only pairs
$\alpha\anti\beta^c, \anti\alpha\beta^c\quad (c=r,y,b)$ cannot be
identified with the FB of SM. Denote
$$
X^c=\anti\alpha\beta^c, \qquad c=r,y,b.
$$
Boson $X$ possesses colors $r,y,b$, electric charge $+2/3$, barion number
$B=1/3$, lepton number $L=-1$. Let's call it a lepto-quark boson.
The consideration of a scheme of gauge field theory, that will be done
in sections 4,5, gives us a possibility to predict the existence of one new
neutral boson
$$
Y={1\over2}(\beta\anti\beta-\sqrt3 \alpha\anti\alpha).
$$
Bosons $X,Y$ have spin 1 and are not yet discovered experimentally. So we may
suggest that their masses are larger then masses of $W^\pm,Z^0$ bosons.
Let us assume the following interactions with the participation of $X$ boson:
$$
\eqalign{
u + \anti\nu_e &\lra X \qquad d+e^+ \lra X\cr
c + \anti\nu_\mu &\lra X \qquad s+\mu^+ \lra X\cr
t + \anti\nu_\tau &\lra X \qquad b+\tau^+ \lra X}
$$
$$
\eqalign{
g^{c\anti c^\prime}+X^{c^\prime}&\lra X^c \quad (c,c^\prime=r,y,b)\cr
\gamma+X&\lra X\cr
Z^0+X&\lra X}
$$
The properties of $Y$ boson are similar to the properties of $Z^0$ boson
- all the FF have nonzero probabilities to emit or absorb boson $Y$.

Strong and electroweak interactions of particles with the participation
of FB and interactions of particles with  the participation of $X,Y$ bosons
can be interpreted as an exchange of constituent parts (inds) of
interacting particles. FB and $X,Y$ are particles that accomplish such an
exchange (there is an evident analogy with the exchange of quarks between
nucleons with the aid of pions). For example, let us consider several
reactions and their interpretations on inds level
$$
\eqalign{
u^r + g^{\anti r y} \lra u^y \quad &\equiv \quad
\beta^r\eps^u\delta^1 + \anti\beta^r\beta^y \lra \beta^y\eps^u\delta^1 \cr
g^{r\anti b}+g^{b\anti y} \lra g^{r\anti y} \quad &\equiv \quad
\beta^r\anti\beta^b+\beta^b\anti\beta^y \lra \beta^r\anti\beta^y \cr
e^- + \anti\nu_e \lra W^-  \quad &\equiv \quad
\alpha\eps^d\delta^1 + \anti\alpha\anti\eps^u\anti\delta^1 \lra
\eps^d\anti\eps^u \cr
d + \anti u \lra W^-  \quad &\equiv \quad
\beta\eps^d\delta^1 + \anti\beta\anti\eps^u\anti\delta^1 \lra
\eps^d\anti\eps^u \cr
u + \anti\nu_e \lra X \quad &\equiv \quad
\beta\eps^u\delta^1 + \anti\alpha\anti\eps^u\anti\delta^1 \lra
\beta\anti\alpha \cr
d+ e^+ \lra X \quad &\equiv \quad
\beta\eps^d\delta^1 + \anti\alpha\anti\eps^d\anti\delta^1 \lra
\beta\anti\alpha \cr
g^{r\anti y} + X^y \lra X^r \quad &\equiv \quad
\beta^r\anti\beta^y + \beta^y\anti\alpha \lra \beta^r\anti\alpha}
$$
In all reactions barion and lepton numbers are conserved. Hence, a decay
of proton is impossible.

Note. Although the interpretation of FB as pairs ind-antiind (or
superpositions of pairs) does not reflect peculiarity of weak interactions
that is connected with the definite handedness of interacting particles, but
it gives us the obvious explanation to all known interactions of
particles and predicts some new interactions. Also, as we shall see in a
moment, this interpretation predetermines the lagrangian which describes
a dynamics of interacting particles. When a lagrangian will be written, FB
may be considered as the corresponding gauge fields.
\bigbreak
{\bf Aquacolor.} There is an evident analogy between present model,
which consider quarks and leptons as trios of inds, and quantum
chromodynamics which consider barions as trios of quarks. It leads to the
assumption that all inds have one more quantum number with three
possible values, which we shall call aquacolor. Antiinds have
antiaquacolor. Suppose, that quarks and leptons are aquacolor singlets
(states antisymmetric in aquacolor). Suppose also an existence of
aquacolor gluons - aquagluons, which are quanta of an aquacolor
interaction that confines trios of inds in quarks and leptons with the
dimension, probably, less then $10^{-17}cm$. It seems natural to
describe aquacolor interaction as a gauge field theory with $SU(3)$
symmetry that corresponds to aquacolor and with the lagrangian (lagrangian
density)
$$
\anti\psi( i \gamma_\mu D_\mu -m) \psi -
{1 \over 4} F_{\mu\nu}^l F_{\mu\nu}^l \eqno(4)
$$
where $\psi$ - 12 component spinor wave function of ind,
$$
D_\mu = \partial_\mu - i a A_\mu, \quad A_\mu = A_\mu^l {\lambda_l \over2}
\quad l=1, \ldots, 8
$$
$$
F_{\mu\nu} = \partial_\mu A_\nu - \partial_\nu A_\mu -
i a [A_\mu A_\nu - A_\nu A_\mu] = F_{\mu\nu}^l {\lambda_l \over2}
$$
$\lambda_l$ -- Gell-Mann's matrices, $\gamma_\mu$ -- Dirac's matrices,
$m$ is not zero only for $\delta^1,\delta^2,\delta^3$.

The lagrangian (3) differs from the QCD lagrangian by
the coupling constant $a$, which must guarantee
a confinement radius less then $10^{-17}cm$. The lagrangian has to describe
interactions of inds due to exchange of aquagluons
on the distances that are less or comparable with
the confinement radius. For the description of interactions of particles due
to exchange of inds on the distances greater than the confinement radius we
have to use another lagrangian.

\bigbreak
{\bf Strong and electroweak interactions.} For a description of dynamics
of strong and electroweak interactions on the inds level at the distances of
order $10^{-16}-10^{-13}cm$ without taking into account interactions
with $X,Y$ bosons, we suggest to use the lagrangian that is immediately
written using rules of SM
$$
\eqalign{
&\anti\Psi i \gamma_\mu(\partial_\mu-
({1\over3}-2\qq){i g_1\over2} B_\mu-{i g_3\over2}\lambda^l C^l_\mu)\Psi \cr
&+\anti \Phi i\gamma_\mu(\partial_\mu-(-1-2\qq){i g_1\over2} B_\mu)\Phi \cr
&+\anti L i \gamma_\mu (\partial_\mu-(2\qq){i g_1\over 2}B_\mu-
{i g_2\over2}\tau^k W^k_\mu)L \cr
&+\anti R^u i\gamma_\mu(\partial_\mu-(1+2\qq){i g_1\over2} B_\mu)R^u \cr
&+\anti R^d i\gamma_\mu(\partial_\mu-(-1+2\qq){i g_1\over2} B_\mu)R^d \cr
&+\sum_{k=1}^3 \anti\Omega_k(i\gamma_\mu\partial_\mu-m_k)\Omega_k\cr
&-{1\over4}B_{\mu\nu}B_{\mu\nu}-{1\over4}W^k_{\mu\nu}W^k_{\mu\nu}
-{1\over4}C^l_{\mu\nu}C^l_{\mu\nu}\cr
&+{|D_\mu \phi|}^2-{1\over2}\lambda^2(|\phi|^2-{1\over2}\eta^2)^2
} \eqno{(5)}
$$
where we have sum over  $\mu,\nu=0,1,2,3$; $k=1,2,3$;
$l=1,\ldots,8$.
$\phi=\pmatrix{\phi_+ \cr \phi_0}$ -- complex dublet of scalar fields
with $Y^w=1$ and the covariant derivative has the form
$$
D_\mu=\partial_\mu - {i g_1\over2}B_\mu - {i g_2\over2}\tau^k W^k_\mu;
$$
$\lambda,\eta$ -- real constants. After the gauge transformation
the dublet $\phi$
has a form
$$
\phi={1\over{\sqrt2}}\pmatrix{0\cr \eta + \chi(x)},
$$
where constant $\eta$ is a vacuum average and $\chi$ -- real scalar
field.  $\Psi,\Phi,L,R^u,R^d,\Omega_k$ -- spinor wave
functions of the following
multiplets of inds:
$$
\Psi=\pmatrix{\beta^r\cr \beta^y\cr \beta^b},\
\Phi=\pmatrix{\alpha},\
L=\pmatrix{\eps^u_L\cr \eps^d_L},\
R^u=\pmatrix{\eps^u_R}, \
R^d=\pmatrix{\eps^d_R},\
\Omega_k=(\delta^k),\ k=1,2,3.
$$
$\anti\Psi, \anti\Phi,\anti L,\anti R^u,\anti R^d, \anti\Omega_k$ -- spinorial
conjugated wave functions. $g_1,g_2$ -- electroweak coupling constants
(sometimes they are denoted by  $g^\prime, g$).
$g_3$ -- strong coupling constant
$$
\eqalign{
        B_{\mu\nu} &= \partial_\mu B_\nu - \partial_\nu B_\mu,\cr
        W_{\mu\nu} & = \partial_\mu W_\nu - \partial_\nu W_\mu -
           i g_2 [W_\mu, W_\nu] = W_{\mu\nu}^k {\tau^k\over2},
           \quad W_\mu=W_\mu^k{\tau^k\over2} \cr
        C_{\mu\nu} & = \partial_\mu C_\nu - \partial_\nu C_\mu -
           i g_3 [C_\mu, C_\nu] = C_{\mu\nu}^l {\lambda^l\over2},
           \quad C_\mu=C_\mu^l{\lambda^l\over2}
}
$$
Values  $C_\mu^l$ define vector fields of gluons. Vector field
$A_\mu$ of photon $\gamma$ and vector fields
 $Z_\mu^0,W_\mu^\pm$ of $Z^0, W^\pm$ bosons calculated as
$$
\eqalign{
A_\mu &={1\over{\sqrt{ {g_1}^2 + {g_2}^2 }}}(g_2 B_\mu + g_1 W_\mu^3)\cr
Z_\mu^0 &={1\over{\sqrt{ {g_1}^2 + {g_2}^2 }}}(- g_1 B_\mu + g_2 W_\mu^3)\cr
W_\mu^\pm &= {1\over\sqrt{2}}(W_\mu^1 \mp i W_\mu^2)
}
$$
A term  $|D_\mu \phi|^2$ gives mass terms for the intermediate bosons
(indices $\mu$ not written)
$$
{\eta^2\over8}(g_2 W^3-g_1 B)^2+{\eta^2 {g_2}^2\over2}W^- W^+
$$
and masses
$$
m_Z={\eta \sqrt{{g_1}^2+{g_2}^2}\over2}, \quad m_W={\eta g_2\over2},
\quad {m_W\over{m_Z}}=\cos{\theta_W}.
$$
Usual formulas connect constants $g_1,g_2$ with the electromagnetic
coupling constant $e$ and Weinberg angle $\theta_W$
$$
e = {g_1 g_2\over \sqrt{ {g_1}^2 + {g_2}^2} } = g_2 \sin{\theta_W},
\qquad \tan{\theta_W} = {g_1\over g_2}.
$$
A mass of scalar field $\chi$ (Higgs particle $H$) has a form
$m_H=\lambda\eta$. Connection between vacuum average $\eta$ and Fermi
constant $G$ is given by formula
$\eta=(\sqrt2 G)^{-1/2}$.

So, the lagrangian (5) describes interactions between spinor fields of inds
$\alpha,\beta, \eps$ and gauge vector fields $B_\mu,C_\mu^l,W_\mu^k$
which define vector massless fields $A_\mu,C_\mu^l$. Fields $A_\mu,C_\mu^l$
can be identified with the physical fields of photon and gluons. In
order to identify fields $Z_\mu^0, W_\mu^\pm$ with the physical fields
of $Z^0,W^\pm$ bosons we have to use a mass generation mechanism for
 $Z^0,W^\pm$ bosons.

In physics there are two well known mass generation mechanisms for
 $Z^0,W^\pm$ bosons that can be included to the present model. The first is
Higgs mechanism that is based on dublet of complex scalar fields
 $(\phi^+,\phi^0)$, which (together with antidublet) have been "eaten"
by $Z_\mu^0, W_\mu^\pm$ fields with the creation of massive $Z^0,W^\pm$
bosons and a scalar Higgs particle. This mechanism is accepted in the SM.
An existence of fundamental scalars in the theory creates some
difficulties (a mass hierarchy problem [20]), which stimulate  development
of models with composite scalar particles. Such models are called
technicolor models [21-24]. In those models the existence is assumed
of new particles -- techniquarks and technigluons which constitute a
composite particles with a confinement radius of order $10^{-17}cm$.
In the second mass generation mechanism bosons $Z,W$ become massive
"eating" goldstone technipions, which appear after spontaneous breaking
of chiral symmetry in quantum technichromodynamics. As was shown by
Weinberg [21], both mass generation mechanisms lead to the same
empirically successful identity $m_W/m_Z=\cos{\theta_W}$.
The evident possibility to identify techniquarks with inds and
technicolor with aquacolor shows that second mass generation mechanism
for $W^\pm,Z^0$ is rather attractive for the use in present model.
\bigbreak

{\bf 5. Inclusion of $X,Y$ bosons.} In order to describe interactions
with the participation of $X,Y$ bosons let us join inds $\beta$ and $\alpha$
in one multiplet. Now $\Psi$ is a column vector which consists of wave
functions of inds $\beta^r,\beta^y,\beta^b,\alpha$. In that case first
two terms in (5) must be replaced by
$$
\anti\Psi i \gamma_\mu(\partial_\mu-
(-2\qq){i g_1\over2} B_\mu-{i g_3\over2}\xi^l C^l_\mu)\Psi+
|D_\mu^\prime \phi_\dd|^2 -
 {1\over2}{\lambda_\dd}^2(|\phi_\dd|^2-{1\over2}{\eta_\dd}^2)^2,
\eqno{(6)}
$$
where $\lambda_\dd, \eta_\dd$ -- real constants,
$\phi_\dd$ --quartet of complex scalar fields with zero weak hypercharge
$Y^w=0$.
$$
\phi_\dd=\pmatrix{\phi_1\cr\phi_2\cr\phi_3\cr\phi_4},\quad
D_\mu^\prime=\partial_\mu-{i g_3\over2}\xi^l C_\mu^l.
$$
The index $l$ has a value from 1 to 15, $\xi^l$ -- traceless
$4\times4$ matrices which are generators of group $SU(4)$
and satisfy a condition ${\sl Tr}{\xi^i\xi^j}=2\delta^{ij}$.
Let us choose
matrices $\xi^l$ in such a way that matrices $\xi^1,\ldots,\xi^8$
have $3\times3$ blocks in upper left corner equal to Gell-Mann's
matrices  $\lambda^1,\ldots,\lambda^8$. The rest set of matrices
$\xi^9,\ldots,\xi^{15}$ we choose in a form
$$
\xi^9=\pmatrix{.&.&.&1\cr.&.&.&.\cr.&.&.&.\cr1&.&.&.},\quad
\xi^{10}=\pmatrix{.&.&.&-i\cr.&.&.&.\cr.&.&.&.\cr i&.&.&.},
$$
$$
\xi^{11}=\pmatrix{.&.&.&.\cr.&.&.&1\cr.&.&.&.\cr.&1&.&.},  \quad
\xi^{12}=\pmatrix{.&.&.&.\cr.&.&.&-i\cr.&.&.&.\cr.&i&.&.},
$$
$$
\xi^{13}=\pmatrix{.&.&.&.\cr.&.&.&.\cr.&.&.&1\cr.&.&1&.},\
\xi^{14}=\pmatrix{.&.&.&.\cr.&.&.&.\cr.&.&.&-i\cr.&.&i&.},\
\xi^{15}={1\over\sqrt{6}}\pmatrix{1&.&.&.\cr.&1&.&.\cr.&.&1&.\cr.&.&.&-3},
$$
where points stand on the places of zeros. Bosons that correspond to
fields $C_\mu^1,\ldots,C_\mu^8$ can be identified with gluons of QCD,
and fields $C_\mu^9,\ldots,C_\mu^{15}$ -- with vector bosons
$X,\anti X, Y$
$$
\eqalign{
X_\mu^r &= {1\over\sqrt2}(C_\mu^{ 9}-i C_\mu^{10}), \quad
\anti X_\mu^r = {1\over\sqrt2}(C_\mu^{ 9}+i C_\mu^{10}), \cr
X_\mu^y &= {1\over\sqrt2}(C_\mu^{11}-i C_\mu^{12}), \quad
\anti X_\mu^y = {1\over\sqrt2}(C_\mu^{11}+i C_\mu^{12}), \cr
X_\mu^b &= {1\over\sqrt2}(C_\mu^{13}-i C_\mu^{14}), \quad
\anti X_\mu^b = {1\over\sqrt2}(C_\mu^{13}+i C_\mu^{14}), \cr
Y_\mu &= C_\mu^{15}.
}
$$
The inclusion of a quartet of scalar fields $\phi_\dd$ with nonzero vacuum
average  $\eta_\dd$ spontaneously breaks $SU(4)$ symmetry of the resulting
lagrangian to color $SU(3)$ symmetry. With the aid of gauge $SU(4)$
transformation the quartet $\phi_\dd$ transforms into
$$
\phi_\dd={1\over\sqrt2}\pmatrix{0\cr0\cr0\cr\eta_\dd+\chi_\dd(x)},
$$
where $\chi_\dd$ -- real scalar field. From a term
$|D_\mu^\prime \phi_\dd|^2$ we get the mass terms for $X,Y$ bosons
$$
{{\eta_\dd}^2 {g_3}^2\over4}(2X^r\anti X^r+2X^y\anti X^y+2X^b\anti X^b+
{3\over2}Y^2)
$$
that give the following masses:
$$
m_X={g_3 \eta_\dd\over2},\quad m_Y={g_3 \eta_\dd\over2}\sqrt{3\over2},\quad
 {m_X\over{m_Y}}=\sqrt{2\over3}.
$$
Gluons remain massless. A mass of scalar field $\chi_\dd$
(new Higgs particle $H_\dd$) has a form $m_{H_\dd}=\lambda_\dd \eta_\dd$.

In that scheme of gauge field theory with $U(1)\times SU(2)\times SU(4)$
symmetry one can see the following analogy: photon $\gamma$ corresponds
to gluons $g_1,\ldots,g_8$; $W^\pm$ bosons correspond to
$X,\anti X$ bosons; $Z^0$ boson corresponds to $Y$ boson.

Now we can formulate postulate 2 which may replace postulate $2^\prime$
from section 2.

\post 2. Bosons of the present model identify with a gauge fields of
lagrangian (5) with (6).\par

So, postulates 1,2 give us a model of elementary particles with $SU(4)$
gauge symmetry which contains all known FF and FB of SM and predicts two
new bosons $X,Y$.
\bigbreak

{\bf 6. Questions.} The lagrangian (5),(6) describes the interactions of
particles on a level of inds. In the description of interactions of
particles on the level of composite FF two very important questions
arise:
\item{-}How to reflect a mixing of $d,s,b$ quarks states (a Cabibbo angle)?
\item{-}What is a mass generation mechanism for FF?

\noindent There are other important questions:
\item{-}Why the selection rule formulated in the section 2 works?
\item{-}What is a value of hidden electric charge $Q^\prime$?
\item{-}Is the present model renormalizable?
\item{-}In what experiments $X,Y$ bosons may be discovered?

\bigskip
We thank A.A.Komar for fruitful discussions.
\medskip
The present work supported by the Russian Fund of Fundamental Research, grant
95-0100433.
\vskip 1cm
email:  nikolai@marchuk.mian.su
\vskip 1cm

Nikolai Marchuk

Steklov Mathematical Institute

Gubkina st.8

Moscow 117966

Russia

\vskip 1cm
\centerline{References}
\vskip 1cm
1. N.G.Marchuk, Doklady of Russian Academy of Science 344, 1 (1995).

2. H.Senju, Prog. Theor. Phys. 46 (1971) 550.

3. H.Senju, Prog. Theor. Phys. 51 (1974) 956.

4. H.Senju, Prog. Theor. Phys. 74 (1985) 413.

5. H.Senju, Prog. Theor. Phys. 92 (1994) 127.

6. J.C.Pati, A.Salam and J.Strathdee, Phys. Lett. 59B (1975) 265.

7. H.Terazawa, Y.Chikashige and K.Akama, Phys. Rev. D15 (1977) 480.

8. H.Terazawa, Phys. Rev. D22 (1980) 184.

9. M.A.Shupe, Phys. Lett. 86B (1979) 87.

10. H.Harari, Phys. Lett. 86B (1979) 83.

11. H.Harari and N.Seiberg, Phys. Lett. 98B (1981) 269.

12. H.Harari and N.Seiberg, Phys. Lett. 100B (1981) 41.

13. K.Sugita, Y.Okamoto and M.Sekine, Nuovo Cim. 104, N8 (1994), 1363.

14. J.C.Pati and A.Salam, Phys. Rev. D10 (1974) 275.

15. K.Matumoto, Prog. Theor. Phys. 52 (1974) 1973.

16. O.W.Greenberg, Phys. Rev. Lett. 35 (1975) 1120.

17. Y.Ne'eman, Phys. Lett. 82B (1979) 69.

18. H.Fritzch and G.Mandelbaum, Phys.Lett. 102B (1981) 319.

19. G.'t Hooft, in "Recent Developments in Gauge Theories", Plenum Press (1980).

20. A.Zee, Unity of Forces in the Universe, World Publishing,
   Singapore, 1982.

21. S.Weinberg, Phys. Rev. D19 (1979) 1277.

22. E.Farhi and L.Susskind, Phys. Rept. 74C (1981) 277.

23. R.K.Kaul, Rev. Mod. Phys. v.55 (1983) 449.

24. R.S.Chivukula, H.Georgi and L.Randall, Nuclear Physics B292(1987) 93.

\end